# Lagrangian Description for Particle Interpretations of Quantum Mechanics – Single-Particle Case

Roderick I. Sutherland

Centre for Time, University of Sydney, NSW 2006 Australia

rod.sutherland@sydney.edu.au

A Lagrangian description is presented which can be used in conjunction with particle interpretations of quantum mechanics. A special example of such an interpretation is the well-known Bohm model. The Lagrangian density introduced here also contains a potential for guiding the particle. The advantages of this description are that the field equations and the particle equations of motion can both be deduced from a single Lagrangian density expression and that conservation of energy and momentum are assured. After being developed in a general form, this Lagrangian formulation is then applied to the special case of the Bohm model as an example. It is thereby demonstrated that such a Lagrangian description is compatible with the predictions of quantum mechanics.

## 1. Introduction

This paper is concerned with the case where the underlying reality between measurements in quantum mechanics is assumed to consist of particles having definite trajectories. An example of this case is the Bohm model (e.g., [1,2]), which successfully reproduces the predictions of standard quantum theory. The present work demonstrates that a Lagrangian description can be formulated consistently and usefully for such particle models. In pursuing the question of the physical interpretation of quantum mechanics, it seems appropriate to seek a Lagrangian description[1] since such a formulation is available in all other areas of physics. For example, a Lagrangian formalism is used in quantum field theory to derive the propagators and vertex factors for Feynman diagrams. In the present case, introducing such a formalism immediately and conveniently provides the equations for describing particle motion, field evolution and conservation laws. It also provides a possible explanation of why quantum phenomena exist at all – in this formalism it is natural to assume the existence of a field arising from the particle itself and influencing the particle's motion so that this motion deviates from the classical predictions.

The treatment given here will be relativistic, although a non-relativistic version can be formulated as well. The relativistic case is chosen because it is actually simpler mathematically. The proposed Lagrangian approach has the advantage of being generally applicable to any wave equation of quantum mechanics. For example, the formulation here is consistent with the Dirac equation. For notational simplicity, the units will be chosen such that $\hbar = c = 1$.

[1] The importance of a Lagrangian approach in this context has also been emphasised by Wharton [3,4].

The discussion here will be limited to the single-particle case and it will be assumed that there is no externally applied potential. This paper is a precursor to a forthcoming, more general one in which the many-particle case is treated.

## 2. Test run – the 4-vector potential of classical electrodynamics

As a preliminary step towards identifying an appropriate Lagrangian expression to choose, the usual expression employed in classical electrodynamics will now be examined[2]. It is well-known that the classical description of a charged particle interacting with an electromagnetic field can be summarised by an overall Lagrangian density $\mathcal{L}$ for the field and particle combined[3]. The overall expression can be written in manifestly Lorentz invariant form as follows:

$$\begin{aligned}\mathcal{L} &= \mathcal{L}_{\text{field}} + \mathcal{L}_{\text{particle}} + \mathcal{L}_{\text{interaction}} \\ &= -\tfrac{1}{4}F_{\alpha\beta}F^{\alpha\beta} - \sigma_0 m\,(u_\alpha u^\alpha)^{1/2} - \sigma_0 q\, u_\alpha A^\alpha\end{aligned} \qquad (\alpha,\beta = 0,1,2,3) \qquad (1)$$

Here:

- $A^\alpha$ is the electromagnetic 4-vector potential,
- $F^{\alpha\beta}$ is the electromagnetic field tensor (which can be expressed in terms of $A^\alpha$),
- $u^\alpha$ is the particle's 4-velocity,
- m and q are the particle's rest mass and charge, respectively,
- $\sigma_0$ is the rest density distribution of the particle through space ($\sigma_0$ will involve a delta function),
- the metric tensor has diagonal elements $(+1,-1,-1,-1)$.

---

[2] I am indebted to a referee for pointing out that classical fields should be treated as distributions, rather than regular smooth functions of spacetime variables, and that products of such distributions as they occur here in $\mathcal{L}_{\text{field}}$ in Eq. (1) can be mathematically ill-defined. Extension procedures of these ill-defined distributions are well-known in the mathematical literature (see, e.g., [5]) and if implemented beforehand lead to finite field theories. Since this issue already arises for standard expressions in physics such as Eqs. (1) and (7) in addition to the new equations introduced below, it will be assumed that the new framework is as mathematically consistent as classical mechanics and so the issue need not be pursued further here.

[3] See, e.g., Eq. (8.38) in [6], or Eq. (13.125) in [7], although both these equations are presented in a somewhat different form.

Both the corresponding field equation and the particle's equation of motion can then be found from the usual Euler-Lagrange equations which are generated via small variations in the field value and the worldline, respectively.

An explicit expression for $\sigma_0$ can be found as follows: The particle's "matter density" is all concentrated at one point in space at any time, so this density has the form:

$$\sigma = \delta^3[\mathbf{x} - \mathbf{x}_p(\tau)] \tag{2}$$

where $\mathbf{x}_p$ is the particle's spatial position as a function of proper time $\tau$ and $\mathbf{x}$ is an arbitrary point in space. However, $\sigma$ is not Lorentz invariant (being just one component of a current 4-vector), so the **rest** density $\sigma_0$ is needed instead. It is given by:

$$\sigma_0 = \frac{1}{u^0}\sigma = \frac{1}{u^0}\delta^3[\mathbf{x} - \mathbf{x}_p(\tau)] \tag{3}$$

Here the factor $u^0$ simply describes the increase in density due to Lorentz contraction as the velocity increases.

Note that we cannot simply apply the usual 4-velocity identity $u_\alpha u^\alpha = 1$ and thereby set the factor $(u_\alpha u^\alpha)^{1/2}$ equal to 1 in the Lagrangian density above. This is because, in using this Lorentz invariant formalism, it is necessary to employ the well-known technique of replacing the proper time with an arbitrary parameter while performing the variation process and so postpone the 4-velocity identity until afterwards.

## 3. Scalar Potential

One can also describe the case of a **scalar** potential $\phi$ interacting with a classical particle by constructing an expression similar to Eq. (1). The appropriate Lagrangian density then has the form:

$$\mathcal{L} = \mathcal{L}_{\text{field}} - \sigma_0 m (u_\alpha u^\alpha)^{1/2} - \sigma_0 \phi (u_\alpha u^\alpha)^{1/2} \tag{4}$$

In this scalar case, the corresponding equation of motion for the particle can be deduced from the Euler-Lagrange equations to be:

$$\frac{d}{d\tau}[(m + \phi)u_\nu] = \partial_\nu \phi \tag{5}$$

The reason for mentioning this equation here is simply to highlight the factor $(m + \phi)$, which shows that a scalar field effectively makes a contribution to the particle's rest mass.

Finally, note that it is necessary to include the factor $(u_\alpha u^\alpha)^{1/2}$ in the last term of Eq. (4), since leaving it out leads to an equation of motion which is inconsistent with the identity $u_\alpha u^\alpha = 1$.

## 4. Application to quantum mechanics

The formulation of the quantum mechanical case now proceeds in analogy to the general approach discussed above. The possible equations of motion for any particle become greatly limited as soon as one makes the following two assumptions: (i) the equations should be Lorentz invariant, and (ii) they should be derivable from a Lagrangian. Equations satisfying these assumptions do exist for each of the following cases: a particle interacting with a scalar potential, a 4-vector potential, a tensor potential of rank two, etc. It will be seen below that agreement with quantum mechanics can be achieved by employing a combination of the scalar and 4-vector cases, as follows:

$$\mathcal{L} = \mathcal{L}_{\text{field}} - \sigma_0 m (u_\alpha u^\alpha)^{1/2} - \sigma_0 \phi (u_\alpha u^\alpha)^{1/2} - \sigma_0 u_\alpha A^\alpha \qquad (6)$$

where specific expressions for the scalar potential $\phi$ and the 4-vector potential $A^\alpha$ are yet to be chosen. Here:

- $\mathcal{L}_{\text{field}}$ is the usual Lagrangian density corresponding to the field in question (Dirac, Klein-Gordon, etc.). This term will be a function of the wavefunction $\psi$ and its complex conjugate $\psi^*$.
- $\phi$ and $A^\alpha$ will also be functions of $\psi$ and $\psi^*$.
- $\sigma_0$, m and $u^\alpha$ are the particle's density distribution, rest mass and 4-velocity, as previously.

## 5. Relevant quantum formalism

At this point, some basic aspects of the standard quantum formalism will be noted.

(i) Each wave equation of quantum mechanics has a probability current density expression $j^\alpha$ associated with it. For example, the current density for the Dirac equation is:

$$j^\alpha = \bar{\psi} \gamma^\alpha \psi \qquad (7)$$

where $\gamma^\alpha$ are the Dirac matrices and $\bar{\psi}$ is the adjoint to the Dirac wavefunction $\psi$.

(ii) The current density can always be written in the form:

$$j^\alpha = \rho_0 \bar{u}^\alpha \qquad (8)$$

where $\rho_0$ and $\bar{u}^\alpha$ are the rest density and 4-velocity **of the probability flow**, respectively. Note that these quantities are quite different from the particle rest density $\sigma_0$ and the particle 4-velocity $u^\alpha$. Also note that the quantities $\rho_0$ and $\bar{u}^\alpha$ in (8) are both uniquely determined once $j^\alpha$ is given, as can be seen by applying the identity $\bar{u}_\alpha \bar{u}^\alpha = 1$ as follows:

$$j_\alpha j^\alpha = (\rho_0 \bar{u}_\alpha)(\rho_0 \bar{u}^\alpha) = {\rho_0}^2 , \quad \therefore \ \rho_0 = (j_\alpha j^\alpha)^{1/2} \tag{9}$$

$$\bar{u}^\alpha = \frac{j^\alpha}{\rho_0} = \frac{j^\alpha}{(j_\beta j^\beta)^{1/2}} \tag{10}$$

## 6. Proposed Lagrangian density

The probability current density $j^\alpha$ discussed in the previous section varies with position and time even in the absence of any externally applied potential, i.e., the flow lines are generally curved, not straight. This means that the motion of the particle presumed to underlie this probability flow is non-uniform as well, i.e., the particle seems to behave as if some sort of field is acting on it. Having noted this point, the aim here is to demonstrate that a "particle-field" system of the sort described by the Lagrangian density (6) can reproduce the standard quantum mechanical predictions. A viable model will now be developed by making the following simple choices for the 4-vector and scalar potentials in terms of $j^\alpha$ and $\rho_0$, respectively:

$$A^\alpha \equiv -k\, j^\alpha \tag{11}$$

$$\phi \equiv k \rho_0 - m \tag{12}$$

where k is an unknown constant with dimensions $[ML^3]$. To keep the formalism simple, this constant will tentatively be set equal to one in what follows. Using Eqs. (11) and (12) with $k = 1$, the general Lagrangian density (6) now simplifies to the following specific form:

$$\mathcal{L} = \mathcal{L}_{\text{field}} - \sigma_0\, \rho_0\, (u_\alpha u^\alpha)^{1/2} + \sigma_0\, u_\alpha j^\alpha \tag{13}$$

Here it is seen that $\rho_0$ effectively acts as the particle's rest mass[4].

---

[4] Note that $\rho_0$ is position dependent. This can be seen as follows: $\rho_0$ is a function of $j^\alpha$ via Eq. (9) and then $j^\alpha$ is a function of $\psi$ via, e.g., Eq. (7). Therefore, since $\psi$ is a function of position, $\rho_0$ is also a function of position.

The various equations implied by the Lagrangian density (13) will now be summarised. The first and third terms of expression (13) have a relatively simple dependence on the wavefunction $\psi$ in that they are both linear in each of $\psi$ and $\psi^*$. This makes deriving the implications of (13) straightforward. The resulting model actually describes a more general situation than quantum mechanics, in that it admits many more solutions. The quantum mechanical case will be obtained at the end by adding an extra assumption so that the system becomes limited to just one state out of those available. In order to extract all desired conclusions, however, it is necessary to formulate the full equations first before introducing this extra assumption.

**7. Field equations**

The field equations can be obtained by applying the usual Euler-Lagrange formula [7]:

$$\partial_\alpha \frac{\partial \mathcal{L}}{\partial(\partial_\alpha \psi^*)} - \frac{\partial \mathcal{L}}{\partial \psi^*} = 0 \tag{14}$$

and inserting (13). In each case (Dirac, Klein-Gordon, etc.), this yields the standard wave equation plus a source term. The source term arises from the non-field parts of $\mathcal{L}$. The existence of this term is necessary to allow action and reaction between the particle and field and hence ensure conservation of energy and momentum. For the special limiting case corresponding to quantum mechanics, however, the source term will be found to vanish, thereby restoring agreement with experiment.

As an example, the appropriate Lagrangian density for the Dirac case is[5]:

$$\mathcal{L} = \left[ -i\bar{\psi}\gamma^\alpha \partial_\alpha \psi + m\bar{\psi}\psi \right] - \sigma_0 \rho_0 (u_\alpha u^\alpha)^{1/2} + \sigma_0 u_\alpha j^\alpha \tag{15}$$

where the quantity in the square bracket is the standard expression for $\mathcal{L}_{\text{field}}$ found in text books. Here the adjoint $\bar{\psi}$ is involved, rather than the complex conjugate $\psi^*$. From this Lagrangian density, the modified form of the Dirac equation is then found to be:

$$i\gamma^\alpha \partial_\alpha \psi - m\psi = \sigma_0 \left( u_\alpha - \bar{u}_\alpha \right) \gamma^\alpha \psi \tag{16}$$

where the left hand side comprises the usual terms, whilst the right hand side is the new source term. This term contains the velocity of probability flow $\bar{u}^\alpha$ as defined in Eq. (10).

---

[5] Since, as mentioned in the introduction, the present discussion is limited to the free-space case, $\mathcal{L}_{\text{field}}$ here does not contain any external potential. A term containing such a potential can easily be added, however. For example, the appropriate extra term for including an external 4-vector potential $A^\alpha$ would be of the form $A_\alpha \bar{\psi} \gamma^\alpha \psi$.

For later reference, the source term for any wave equation (i.e., for any $j^{\alpha}$) has the general form:

$$-\partial_{\alpha}\left[\sigma_0\left(u_{\beta}-\bar{u}_{\beta}\right)\frac{\partial j^{\beta}}{\partial(\partial_{\alpha}\psi^*)}\right]+\sigma_0\left(u_{\beta}-\bar{u}_{\beta}\right)\frac{\partial j^{\beta}}{\partial\psi^*} \tag{17}$$

Note that this formulation implies that the source of the mysterious field apparently acting on a particle to produce the quantum predictions is actually the particle itself.

**8. Equations of motion for the particle**

To derive this equation a Lagrangian L is needed, rather than a Lagrangian density $\mathcal{L}$. For this purpose, Eq. (13) can be re-expressed in the form:

$$\mathcal{L}=\mathcal{L}_{\text{field}}+\sigma_0 L \tag{18}$$

where:

$$L=-\rho_0(u_{\alpha}u^{\alpha})^{1/2}+u_{\alpha}j^{\alpha} \tag{19}$$

Inserting this expression into the usual Euler-Lagrange formula for a particle [7]:

$$\frac{d}{d\tau}\frac{\partial L}{\partial u^{\alpha}}=\frac{\partial L}{\partial x^{\alpha}} \tag{20}$$

yields the following general equation of motion:

$$\frac{d}{d\tau}(\rho_0 u_{\alpha})=\partial_{\alpha}\rho_0+u^{\beta}(\partial_{\beta}j_{\alpha}-\partial_{\alpha}j_{\beta}) \tag{21}$$

which can be used in conjunction with any choice of wave equation.

**9. Energy-momentum tensor**

Under the assumption that the Lagrangian density is not an explicit function of the coordinates $x^{\alpha}$ (i.e., symmetric under space and time displacements), Noether's theorem implies the existence of an energy-momentum tensor for the particle-field system, with this tensor having zero divergence:

$$\partial_{\beta}T^{\alpha\beta}=0 \tag{22}$$

This condition ensures overall conservation of energy and momentum. The tensor $T^{\alpha\beta}$ is found to break up naturally into three terms (not separately conserved):

$$T^{\alpha\beta}=T^{\alpha\beta}_{\text{field}}+T^{\alpha\beta}_{\text{particle}}+T^{\alpha\beta}_{\text{int eraction}} \tag{23}$$

A straightforward calculation leads to the following general expressions for these terms[6]:

$$T^{\alpha\beta}_{\text{field}} = (\partial^\alpha \psi)\frac{\partial \mathcal{L}_{\text{field}}}{\partial(\partial_\beta \psi)} + (\partial^\alpha \psi^*)\frac{\partial \mathcal{L}_{\text{field}}}{\partial(\partial_\beta \psi^*)} - g^{\alpha\beta}\mathcal{L}_{\text{field}} \tag{24}$$

$$T^{\alpha\beta}_{\text{particle}} = \sigma_0\,\rho_0\, u^\alpha u^\beta \tag{25}$$

$$T^{\alpha\beta}_{\text{interaction}} = \sigma_0\left(u_\lambda - \bar{u}_\lambda\right)\left\{(\partial^\alpha \psi)\frac{\partial j^\lambda}{\partial(\partial_\beta \psi)} + (\partial^\alpha \psi^*)\frac{\partial j^\lambda}{\partial(\partial_\beta \psi^*)}\right\} - \sigma_0\, j^\alpha u^\beta \tag{26}$$

which can be used with any choice of wave equation. The expression (24) for $T^{\alpha\beta}_{\text{field}}$ is the standard one found in textbooks.

**10. Requirements for a particle model to agree with quantum mechanics**

The requirement of consistency with quantum mechanics imposes statistical restrictions on the position and velocity values that can be possessed by an ensemble[7] of particles. It will be more convenient here to express these restrictions in terms of the 3-velocity $\mathbf{v} = \frac{d\mathbf{x}}{dt}$ rather than the 4-velocity $u$. Note that the particle velocities discussed here correspond to values existing between measurements but not necessarily to measurement outcomes (since the measurement interaction may gradually modify the state during the measurement process, as in the Bohm model [1]).

Now, if we describe the range of position and velocity values existing in an ensemble at time t by a joint probability distribution $\rho(\mathbf{x},\mathbf{v};t)$, a necessary condition for consistency is that the distribution should be related to the relevant quantum mechanical current density $j^\alpha$ in the following two ways:

$$\int_{-\infty}^{+\infty} \rho(\mathbf{x},\mathbf{v})\, d^3v \equiv \rho(\mathbf{x}) = j^0(\mathbf{x}) \tag{27}$$

$$\int_{-\infty}^{+\infty} \rho(\mathbf{x},\mathbf{v})\, v^i\, d^3v \equiv \rho(\mathbf{x})\left\langle v^i \right\rangle_{\mathbf{x}} = j^i(\mathbf{x}) \qquad (i = 1,2,3) \tag{28}$$

---

[6] The overall tensor $T^{\alpha\beta}$ defined here is actually the "canonical" energy-momentum tensor. In the case of spin-zero fields (e.g., the Klein-Gordon case) it is found to be symmetric under interchange of $\alpha$ and $\beta$, which implies conservation of angular momentum as well. In other cases, such as the Dirac field, $T^{\alpha\beta}$ is not symmetric. This is due to the fact that the canonical tensor only provides a description of orbital angular momentum density, not spin angular momentum density. Since these two contributions are not separately conserved, the canonical tensor is not an adequate description. Techniques exist to symmetrise this tensor and so include the spin part as well [8].

[7] "Ensemble" in the present context refers to single particles considered in different runs of the same experiment.

where $\langle v^i \rangle_{\mathbf{x}}$ is the mean value of $v^i$ at point $\mathbf{x}$. The fact that these equations can also be taken as a sufficient condition for consistency is argued in Bohm's theory of measurement [1,2] and will be tentatively accepted here.

**11. General Case**

A variety of different particle models are then possible by choosing different expressions for $\rho(\mathbf{x},\mathbf{v})$. The joint probability distribution chosen will need to be a function of the relevant wavefunction $\psi$ in order to satisfy (27) and (28). Progress is usually impeded at this point by the requirement that the probability density $\rho(\mathbf{x},\mathbf{v})$ should also be positive, which is not easily satisfied. For example, the Wigner distribution [9] is often used for practical calculations, but the fact that it is not positive means that it is not physically viable for present purposes. Note that each model will generally allow a range of different velocity values at each point in space-time (i.e., it will be a "phase space" model). The simplest model, however, is obtained by restricting the particle's velocity to a unique value at each point (i.e., choosing the particle's velocity $\mathbf{v}$ to be a function $\mathbf{v}(\mathbf{x})$ of position). This reduces the description from phase space to configuration space and thereby provides a positive probability distribution. In so doing, it yields the Bohm model.

**12. Special Case - Bohm**

The Bohmian case corresponds to assuming the following joint distribution:

$$\rho(\mathbf{x},\mathbf{v}) = j^0(\mathbf{x})\,\delta^3(\mathbf{v}-\overline{\mathbf{v}}) \tag{29}$$

where $\mathbf{v}$ and $\overline{\mathbf{v}}$ are the 3-velocities of the particle and the probability flow, respectively. For the purposes of the present Lagrangian formulation, we simply need the 4-velocity version of this assumption, which is as follows:

$$u^\alpha = \overline{u}^\alpha \tag{30}$$

Here $u^\alpha$ and $\overline{u}^\alpha$ are the 4-velocities of the particle and the probability flow, as defined earlier[8]. Eq. (30) comprises the extra "quantum consistency condition" to be added to the earlier formalism.

**13. Implications of the quantum consistency condition**

The simplifications this condition produces in the Lagrangian formulation will now be listed.

---

[8] Note that, although the particle velocity is constrained by Eq. (30) to be equal to the local current velocity, the converse is not necessarily true (in the sense that the current velocity is obviously not equal to the particle velocity at locations where the particle is absent).

(i) Field Equation:

Inserting $u^\alpha = \bar{u}^\alpha$ into the general source term (17) is seen to reduce this term to zero, so that only the standard wave equation remains and agreement with experiment is assured.

(ii) Energy-momentum tensor:

The effect of the extra condition (30) in this case is that the term $T^{\alpha\beta}_{interaction}$ becomes simpler and the divergence of $T^{\alpha\beta}_{field}$ becomes separately zero.

(iii) Particle Equation of Motion:

This differential equation would normally allow a range of possible solutions for the particle velocity, depending on the boundary conditions imposed. Here, however, $u^\alpha$ is being restricted to the particular solution $u^\alpha = \bar{u}^\alpha$. It needs to be confirmed that this result is, in fact, a consistent solution to the equation of motion (21). That this is true can be shown by using Eq. (8) to rewrite the relationship $u^\alpha = \bar{u}^\alpha$ in the following equivalent form:

$$j^\alpha = \rho_0 u^\alpha \tag{31}$$

Substituting (31) into (21) is then easily seen to reduce the latter to an identity, as required.

In the Bohmian case it is evident that the extra assumption for consistency with quantum mechanics introduces a considerable simplification in the previously-formulated equations.

## 14. Justification for the extra restriction imposed

It would be preferable to be able to derive the quantum consistency condition directly from some extra variation to be performed on the action function, rather than just postulating it separately. One possibility is to require that the action S also be stationary under small variations in the particle 4-velocity u. Referring back to Eq. (18):

$$\mathcal{L} = \mathcal{L}_{field} + \sigma_0 L$$

and noting that $\mathcal{L}_{field}$ and $\sigma_0$ are not dependent on u, this condition can be expressed as:

$$\delta S = \int \delta L \, d\tau = \int \frac{\partial L}{\partial u^\alpha} \delta u^\alpha \, d\tau = 0 \tag{32}$$

For this to hold for any small $\delta u^\alpha$ requires[9]:

$$\frac{\partial L}{\partial u^\alpha} = 0 \tag{33}$$

Evaluating $\frac{\partial L}{\partial u^\alpha}$ from (19) yields:

$$\begin{aligned}\frac{\partial L}{\partial u^\alpha} &= \frac{\partial}{\partial u^\alpha}\left[-\rho_0 (u_\beta u^\beta)^{1/2} + u_\beta j^\beta\right] \\ &= -\rho_0 (u_\beta u^\beta)^{-1/2}\, u_\alpha + j_\alpha\end{aligned} \tag{34}$$

and setting this equal to zero then yields the equation of motion:

$$u^\alpha = \frac{j^\alpha}{\rho_0} \tag{35}$$

i.e.,

$$u^\alpha = \overline{u}^\alpha \tag{36}$$

Hence this may be an appropriate way to justify the quantum consistency condition.

## 15. Discussion and Conclusions

A Lagrangian formulation has been presented here in the context of the underlying reality between measurements in quantum mechanics being particles with definite trajectories. The formalism is then used to derive expressions for the corresponding particle equation of motion, field equation and energy-momentum tensor, thereby demonstrating the richness which becomes available as soon as this approach is pursued. In particular, it automatically provides a self-interaction picture to explain quantum phenomena.

After formulating the Lagrangian description, the conditions for it to be consistent with the predictions of quantum mechanics have been examined and presented. These take the form of two further restrictions to be placed on the general formalism (Eqs. (27) and (28)). The Bohm model has then been highlighted as a special case where this consistency with quantum mechanics is evident.

An extension of this model to the many-particle case will be presented shortly. Maintaining Lorentz invariance in this more general case, however, will be seen to require the addition of retrocausality, as described previously in [11].

---

[9] A similar condition to that imposed by Eq. (33) has been employed by other authors, e.g., Eq. (44) in [10].)

## Acknowledgements

Thanks are due to Ken Wharton and to David Miller for kindly providing helpful feedback on this work.